\documentclass[twocolumn,secnumarabic,amssymb, nobibnotes, aps, pra,showpacs,preprintnumbers]{revtex4-1}
\usepackage{amssymb}
\usepackage{array}
\usepackage{graphicx}

\def\beq{\begin{equation}}
\def\eeq{\end{equation}}
\def\bea{\begin{eqnarray}}
\def\eea{\end{eqnarray}}
\def\no{\nonumber}

\begin{document}

\title{Non-Abelian Aharonov-Bohm effect with the time-dependent gauge fields}%

\author{Seyed Ali Hosseini Mansoori$^{1,2}$ and Behrouz Mirza$^2$ }
\affiliation{$^1$ Department of Physics, Boston University, 590 Commonwealth Ave., Boston, MA 02215, USA\\
$^2$Department of Physics, Isfahan University of Technology, Isfahan 84156-83111, Iran}
\email{shossein@bu.edu; \,\,\ \\
b.mirza@cc.iut.ac.ir; \,\,\ \\}
\date{\today}%

\begin{abstract}
 We investigate the non-Abelian  Aharonov-Bohm (AB) effect for time-dependent gauge fields. We prove that  the non-Abelian AB phase shift  related to time-dependent gauge fields, in which the electric and magnetic fields are written in the adjoint representation of $SU(N)$ generators, vanishes up to the first order expansion of the phase factor. Therefore, the flux quantization in a superconductor ring does not appear in the time-dependent Abelian or non-Abelian AB effect.

\end{abstract}


\maketitle
%\tableofcontents

\section{Introduction}\label{a}
%The vector potential plays a predominant role in quantum mechanics
%because it appears explicitly in
%the equation of motion. Moreover, it is not directly observable, and might be determined only by a gauge transformation.
In 1959, Y. Aharonov and D. Bohm proposed an experiment to test the  effect of
the electromagnetic gauge potential on the quantum wave function \cite{ah1}.  Later, Chambers performed the proposed experiment and proved that the effect did exist \cite{cham}. The AB effect is indeed a quantum-mechanical phenomenon in which the wave function of a charged particle traveling  around an extremely long solenoid  undergoes a phase shift depending on the magnetic field between the paths  albeit $B = 0$ along the paths themselves  \cite{ah1, cham}.

Over the past few years,  considerable interest has been shown in the AB effect in Abelian gauge fields with a time-independent magnetic field. Recently, the AB effect with a time-dependent magnetic field has been investigated \cite{do1,do2,go1,go2,go4} to show that a cancelation of phases occurs in the AB effect with a time-dependent magnetic field. Strictly speaking, an extra phase  coming from the electric field, $E = -\partial_{t}A$, outside the solenoid  cancels out the phase shift of the time-dependent magnetic field. The experimental results of Marton et al. \cite{go5}, where the effect of the time variation of the magnetic field was not seen in the interference pattern, also confirm the theoritical prediction of \cite{go6}, i.e., an exact cancellation of the AB
phase shift by means of the phase shift coming from the direct Lorentz force. In this framework, the time-dependent AB effect can be considered as a type II AB effect. Indeed, the type I effects are in situations that a charged particle is moving through a region without magnetic and electric fields, while the type II AB effects are when the charged particle develops an AB phase passing through a region of space with non-zero fields \cite{go7}. 
Recently, in Ref. \cite{go8}, authors have shown that type II AB effect due to electromagnetic plane waves vanishes under some conditions in terms of the parameters of the system like frequency of the electromagnetic wave, the size of the space-time loop, and amplitude of the electromagnetic wave. 

%%On the other hand,  in the non-Abelian cases, however, the strength tensor is not gauge invariant. In order to define a gauge-invariant field tensor,  all fields must be %%written in the adjoint representation. Therefore, i
 It is, therefore, interesting to study the non-Abelian AB effect \cite{P1} with a time-dependent  magnetic field. Recently, the AB effect has been studied for time-dependent non-Abelian fields by using two specific, known time-dependent solutions \cite{d1} such as the Coleman plane wave solutions \cite{d2} and the time-dependent Wu-Yang monopole \cite{d3}.  Here, we prove that when the non-Abelian gauge field $A^{a}$ is a function of spacetime, the AB phase shift  coming from the electric and magnetic non-Abelian fields will be canceled out up to the first order. Our results also show that the "single-valuedness of the wave function" does not constrain the flux of a time-dependent magnetic field to be quantized in a superconducting ring. It will be interesting to verify this result experimentally.

The outline of this paper is as follows. Section II presents  a description of the AB phase shift. In Section III, we study the quantization of the magnetic flux in a superconducting ring. We will show that there is no "single-valuedness" condition for the wave function because the phase shift will be zero in a time-dependent AB effect.  In Section IV, we generalize the Abelian AB effect to a time-dependent non-Abelian field configuration. We prove that the AB phase factor remains equal to  zero up to the first order when considering the time-varying vector fields. Conclusions will be presented  in the last Section.

%%%%%%%%%%%%%%%%%%%%%%%%%%%%%%%%%%%%%%%%%%%%%%%%%%%%%%%%%%%%%%%%%%%%%%%%%%%%%%%%%%%%%%%%%%%%%%%%%%%%%%%%%%%%%%%%%%%%%%%%%
\section{Time-dependent AB Effect for Abelian gauge fields }
The relativistic form of the AB phase factor can be written as follows:
\begin{equation}\label{g1}
\beta =exp\left[ \frac{e}{\hbar }\oint {{A}_{\mu }}d{{x}^{\mu }} \right]=exp\left[ \frac{e}{\hbar }\oint{\varphi dt-A.dx} \right]
\end{equation}
where, $A^{\mu}$ is the Abelian gauge field that might be transformed under the $U(1)$ group as follows:
\begin{equation}
{{A}^{\mu }}\to {{A}^{\mu }}^{\prime }={{A}^{\mu }}+{{\partial }^{\mu }}\xi
\end{equation}
where, $\xi$ is a transformation function of space-time coordinates \cite{kh,ry1}.
We may rewrite  the phase factor in a 2-form structure by making use of Stokes' theorem, stating that the integral of a differential form $\omega$ over the boundary of some orientable manifold $\Omega$ is equal to the integral of its exterior derivative $d\omega$ over the whole of $\Omega$, which may be expressed as follows:
 \begin{equation}\label{s2}
\int_{\partial \Omega}\omega=\int_{\Omega}\mathrm {d}\omega
 \end{equation}
 where, $\omega$ and $d\omega$ are p-form and $(p + 1)$-form, respectively.
One could also define the 1-form as $\omega =A={{A}_{\mu }}d{{x}^{\mu }}$ and 2-form $d\omega=dA$ as the Faraday 2-form $F$ by:
\begin{eqnarray}\label{g3}
&dA=F=\frac{1}{2}{{F}_{\mu \nu }}d{{x}^{\mu }}\wedge d{{x}^{\nu }}= \\
\nonumber & ({{E}_{x}}dx+{{E}_{y}}dy+{{E}_{z}}dz)\wedge dt+{{B}_{x}}dy\wedge dz\\
&\no +{{B}_{y}}dz\wedge dx+{{B}_{z}}dx\wedge dy
\end{eqnarray}
where $E$ and $B$ are the electric  and magnetic fields, respectively. Therefore,  Eq. (\ref{g1}) can be rewritten as  in (\ref{g2}) below:
\begin{equation}\label{g2}
\beta=exp\left[-\frac{e}{2\hbar }\int{{{F}_{\mu \nu }}d{{x}^{\mu }}\wedge d{{x}^{\nu }}}\right]
\end{equation}
This expression plays a key role in the study of the AB phase factor when  considering the time-dependent Abelian gauge fields.
Time-dependent AB Effect is based on constructing a subspace in a spacetime
in which the four-vector potential depends on time \cite{bl1, bl2}. Both the electric and the magnetic effects depend on
the particle's particular path in this subspace \cite{do2}. We assume that  the magnetic field inside the solenoid is time-dependent so that the vector potential $A$ will be time-dependent outside the solenoid. However, based on Maxwell's equation, i.e.,  $E = -\partial_{t}A$, an electric field is also created outside the solenoid (We have assumed the scalar potential field $\varphi$ to be zero).   Thus, from Eqs. (\ref{g3}) and (\ref{g2}), the magnetic phase factor is obtained by:
\begin{eqnarray}
&\frac{e}{\hbar }\int{\left[ {{B}_{x}}dy\wedge dz+{{B}_{y}}dz\wedge dx+{{B}_{z}}dx\wedge dy \right]}\\
\no&=\frac{e}{\hbar }\int{B(x,t).dS}
\end{eqnarray}
and the electric part of the phase is given by:
\begin{eqnarray}
&\frac{e}{\hbar }\int{\left[ {{E}_{x}}dx\wedge dt+{{E}_{y}}dy\wedge dt+{{E}_{z}}dz\wedge dt \right]}\\
\no&=-\frac{e}{\hbar }\oint{A.dx=}-\frac{e}{\hbar }\int{B(x,t).dS}
\end{eqnarray}
where, we have replaced the electric field by $-\partial_{t}A$.
It is clear that the AB phase shift for a
time-varying magnetic field vanishes. This means that the magnetic AB
phase shift is canceled out by a phase shift coming from the Lorentz force associated with the electric field,
$E = -\partial_{t}A$, outside the solenoid \cite{do1}.

\section{Non-flux quantization in superconducting rings for time-dependent magnetic fields }
Let us now consider a superconducting ring with rigid walls which is exposed to an external uniform magnetic field.  Assuming a particle of charge $e$ completely confined in the interior shell of the superconducting ring, one can obtain the relevant energy eigenvalues and wave functions \cite{chi, da, y1}.
 However, care must be taken to ensure  that the value of the wave function at any given point in the ring has the same value as the wave function obtained by traveling around the ring to return back to the original point. In other words,  the wave function must have a single value at a given point in the ring.

%The single-valuedness of the wave function constrains the flux to be quantized values.  According to electromagnetic theory, applying a magnetic field to a superconductor induces a change in the phase of the wave function.  In the other words, a given amount of magnetic field creates a specific phase change in the wave function as if this phase depends upon the applied magnetic field.
In this case, the variation of the wave function phase is,
\begin{eqnarray}\label{j2}
&\delta \alpha_{B} =\frac{2e}{\hbar }\oint{A.dl}=\frac{2e}{\hbar }\int{\nabla \times A}.dS\\
\no&=\frac{2e}{\hbar }\int{B.dS}=\frac{2e}{\hbar }\Phi
\end{eqnarray}
where, $\Phi$ is the magnetic flux and the factor 2e shows that the Cooper pairs \cite{misner} in the superconductor have  charges twice that of an  electron. In order to maintain the single-valuedness of the wave function, this phase factor must be equal to $2\pi n$ ( $n=1,2,3,...$), so that
%\begin{equation}\label{gg4}
%\delta {{\alpha }_{B}}=\frac{2e\Phi }{\hbar }=2\pi n \,\,\,\;\,\,\,\ n=1,2,3,...
%\end{equation}
we can obtain the following quantum  flux,
\begin{equation}
{{\Phi }_{n}}=\frac{\hbar \pi n}{e}=\frac{hn}{2e} \,\,\,\,\;\,\,\,\,\ n=1,2,3,...
\end{equation}
Now, we consider a time-dependent magnetic field. According to Maxwell's equation, there is an electric field ( $E=-\nabla \varphi -{{\partial }_{t}}A$) which creates an additional phase factor. Moreover, for this case, the scalar potential is still zero and the vector potential is a function of time and space. Based on Eq. (\ref{g3}), the relativistic phase shift will be zero due to the cancelation of the magnetic phase shift due to a phase shift coming from the electric field,
$E = -\partial_{t}A$.
 As a result, there is no constraint  on the magnetic flux $\Phi$. It will be interesting  to design an experimental plan  to examine this effect.
\section{Time-dependent AB Effect for non-Abelian gauge fields }
  In section II, we investigated the time-dependent AB effect  \cite{do1,go1,go2}, and showed that there is no  phase shift in this case. In this section, we will verify the claim that  the phase shift of the non-Abelian AB effect is zero for time-dependent  gauge fields.

The concept of the non-Abelian gauge field was first introduced  in 1954 by Yang and Mills \cite{chen}.
 The 4-vector gauge fields $A_{a}$ were introduced with N internal components
labeled by $a = 1, 2, 3,...,N$, corresponding to the N-generators of the
gauge group closed under the following commutation;
\begin{equation}\label{p1}
\left[ {{L}_{a}},{{L}_{b}} \right]=iC_{ab}^{c}{{L}_{c}}
\end{equation}
where, the constants $C_{ab}^{c}$ are real numbers called structure constants.
For simplicity, we shall use the shorthand ${{A}^{\mu }}={{A}^{\mu }}_{a}{{L}_{a}}$, which is a
matrix. Moreover, ${{A}^{\mu }}_{a}$ under an infinitesimally local gauge transformation can be written
in the following form,
\begin{equation}
A_{a}^{\mu }(x)\to A_{a}^{\mu }(x)+\frac{1}{g}{{\partial }^{\mu }}{{\omega }_{a}}(x)+{{C}_{abc}}{{\omega }_{b}}(x)A_{c}^{\mu }(x)
\end{equation}
where, $g$ and $\omega$ are the gauge coupling constant and arbitrary real functions, respectively. In Maxwell's $U(1)$ gauge theory, a gauge-invariant field tensor ${{F}_{\mu \nu }}={{\partial }_{\mu }}{{A}_{\nu }}-{{\partial }_{\nu }}{{A}_{\mu }}$ is defined, whose components are the electric and magnetic fields; in the non-Abelian case, however, such a field tensor is not gauge invariant or, indeed, there is no gauge-invariant field tensor.
In order to define a gauge-invariant field tensor for the non-Abelian gauge fields, the representation must be the adjoint representation \cite{mn1}. The following equation satisfies our requirements,
\begin{equation}\label{s4}
F_{\mu \nu }^{a}={{\partial }_{\mu }}A_{\nu }^{a}-{{\partial }_{\nu }}A_{\mu }^{a}+g{{C}^{abc}}A_{\mu }^{b}A_{\nu }^{c}
\end{equation}
 where, the antisymmetric constants $C_{abc}=-i(L_{b})_{ac}$ are defined in the adjoint representation.
 Using the above equation, the electric and the magnetic fields can be written as in the following equations \cite{kh},
 \begin{equation}\label{s1}
 {{E}_{a}}=-\nabla A_{a}^{0}-{{\partial }_{t}}{{A}_{a}}-g{{C}_{{abc}}}{{A}_{b}}A_{c}^{0}
 \end{equation}
 \beq\label{s8}
{{B}_{a}}=\nabla \times {{A}_{{a}}}+\frac{1}{2}g{{C}_{abc}}{{A}_{{b}}}\times {{A}_{{c}}}
\eeq
It is surprising that the AB experiment can also be used to examine the existence of non-Abelian gauge fields \cite {f1, TT}.
 One can generalize the phase factor of the Abelian AB effect to the non-Abelian AB one \cite{P1} using the following Relation:
\begin{equation}\label{k1}
\beta =P\left[ \exp \left[ \frac{g}{\hbar }\oint{A} \right] \right]
\end{equation}
where, $A=A_{\mu }^{a}{{L}_{a}}d{{x}^{\mu }}$ and $P$ is the path-ordering operator. This phase factor is quite similar to Wilson loop \cite{kkk1,go9}.  Expanding this phase shift up to the second order, we will have:
\begin{eqnarray}\label{f1}
&\beta \simeq 1+\frac{g}{\hbar }{{L}_{a}}\oint{A_{\mu }^{a}d{{x}^{\mu }}}+\\
& \nonumber P(\frac{g}{\hbar })^2\oint{\oint{d{{x}^{\mu }}{{A}^{a}}_{\mu }(x)d{{x}^{\nu }}.{{A}_{\nu }}^{b}(x)}}{{L}_{a}}{{L}_{b}}+...
\end{eqnarray}
We will now go on  to show that the time-dependent non-Abelian AB phase shift vanishes up to the first order, while the other orders indicate a non-zero non-Abelian AB phase factor.
Let us consider a 4-vector potential in the non-Abelian AB effect as $A_{a}^{\mu }\equiv (A_{a}^{0},A_{a}^{i})=(0,A_{a}^{i}(x,t))$. Therefore, from Eq. (\ref{s1}), the electric field will be a non-zero  term (${{E}_{a}}=-{{\partial }_{t}}{{A}_{a}}$).
Applying the Stoke's theorem (Eq. \ref{s2}), we will have:
\begin{equation}
d\omega =dA=dA^{a}L_{a}=\left( {{\partial }_{\mu }}A_{\nu }^{a}-{{\partial }_{\nu }}A_{\mu }^{a} \right){{L}_{a}}d{{x}^{\mu }}\wedge d{{x}^{\nu }}
\end{equation}
Based on Eq.  (\ref{s4}), the above equation can be rewritten as:
\begin{eqnarray}\label{s9}
d{{A}^{a}}=\frac{1}{2}F_{\mu \nu }^{a}d{{x}^{\mu }}\wedge d{{x}^{\nu }}-g{{C}^{abc}}A_{\mu }^{b}A_{\nu }^{c}d{{x}^{\mu }}\wedge d{{x}^{\nu }}
\end{eqnarray}
where, the factor  $1/2$ comes from the anti-symmetry property of $F_{\mu\nu}$ and $dx^{\mu}\wedge dx^{\nu}$ \cite{mn1}.
Therefore, the second term of the expansion in Eq. (\ref{f1}) may be replaced with the following equation:
\begin{eqnarray}\label{f2}
\no&\frac{g}{\hbar }{{L}_{a}}\oint{A_{\mu }^{a}d{{x}^{\mu }}}=\frac{g}{2\hbar }{{L}_{a}}\int{F_{\mu \nu }^{a}d{{x}^{\mu }}\wedge d{{x}^{\nu }}}\\
&-\frac{{{g}^{2}}}{\hbar }{{C}^{abc}}{{L}_{a}}\int{A_{\mu }^{b}A_{\nu }^{c}d{{x}^{\mu }}\wedge d{{x}^{\nu }}}
\end{eqnarray}
where, the 2-form tensor can be defined as,
\begin{eqnarray}\label{s6}
&\no \frac{1}{2}{{F}_{\mu \nu }}^{a}d{{x}^{\mu }}\wedge d{{x}^{\nu }}= ({{E^{a}}_{x}}dx+{{E^{a}}_{y}}dy+{{E^{a}}_{z}}dz)\wedge dt\\
&+{{B^{a}}_{x}}dy\wedge dz +{{B^{a}}_{y}}dz\wedge dx+{{B_{a}}_{z}}dx\wedge dy
\end{eqnarray}
One can then divide up  the above equation into the two  magnetic and electric parts. In this way,  the phase difference associated with the magnetic field terms is given by:
\begin{eqnarray}
&\frac{g}{\hbar }\int{\left[ B_{x}^{a}dy\wedge dz+B_{y}^{a}dz\wedge dx+B_{z}^{a}dx\wedge dy \right]}=\\
\no&\frac{g}{\hbar }\int{{{B}^{a}}(x,t).dS}
\end{eqnarray}
Substituting $B^{a}$ from Eq. (\ref{s8}) and using the Stoke's theorem, we have:
\begin{equation}\label{s12}
\frac{g}{\hbar }\int{{{B}^{a}}(x,t).dS}=\frac{g}{\hbar }\oint{{{A}^{a}}.dl+\frac{g^2}{2\hbar }{{C}^{abc}}\int{({{A}^{b}}\times {{A}^{c}}).dS}}
\end{equation}
On the other hand, the electric field part is given by:
\begin{eqnarray}\label{s13}
&\frac{g}{\hbar }\int{\left[ E_{x}^{a}dx\wedge dt+E_{y}^{a}dy\wedge dt+E_{z}^{a}dz\wedge dt \right]}\\
\no&=-\frac{g}{\hbar }\oint{{{A}^{a}}.dl}
\end{eqnarray}
where, the electric field is ${{E}_{a}}=-{{\partial }_{t}}{{A}_{a}}$. Finally, one can obtain the phase shift from the 2-form tensor as follows:
\begin{equation}\label{f3}
\frac{g}{2\hbar }{{L}_{a}}\int{F_{\mu \nu }^{a}d{{x}^{\mu }}\wedge d{{x}^{\nu }}}=+\frac{g^2}{2\hbar }{{C}^{abc}}L_{a}\int{({{A}^{b}}\times {{A}^{c}}).dS}
\end{equation}
Moreover, when considering the 4-vector potential  $A_{a}^{\mu }\equiv (A_{a}^{0},A_{a}^{i})=(0,A_{a}^{i}(x,t))$, we can rewrite the second part of Relation (\ref{f2}) as follows:
\begin{eqnarray}\label{s10}
\no &-\frac{{{g}^{2}}}{\hbar }{{C}^{abc}}{{L}_{a}}\int{A_{\mu }^{b}A_{\nu }^{c}d{{x}^{\mu }}\wedge d{{x}^{\nu }}}\\
 \no &=-\frac{{{g}^{2}}}{\hbar }{{C}^{abc}}{{L}_{a}}\int{(A_{j}^{b})(A_{k}^{c})}d{{x}^{j}}\wedge d{{x}^{k}}\\
&=-\frac{g^2}{2\hbar }{{C}^{abc}}L_{a}\int{({{A}^{b}}\times {{A}^{c}}).dS}
\end{eqnarray}
 in which the following wedge product is used
  \begin{equation}
d{{x}^{j}}\wedge d{{x}^{k}}={\frac{{\varepsilon }^{ijk}}{2}}d{{x}^{j}}\otimes d{{x}^{k}}
\equiv \frac{{\varepsilon }^{ijk}}{2}dS_{i}.
\end{equation}
Finally, using Eqs.  (\ref{f2}), (\ref{f3}), and (\ref{s10}), we arrive at the following interesting result:
\begin{equation}
\frac{g}{\hbar }{{L}_{a}}\oint{A_{\mu }^{a}d{{x}^{\mu }}}=0
\end{equation}
 Therefore, the phase shift related to the time-dependent  non-Abelian  AB effect vanishes up to the first order  expansion of the phase factor. This is a generally valid result. For future research, it will be  interesting to investigate the higher order terms of gauge fields. It may be anticipated  that all higher order terms of gauge fields will also vanish.  This conjecture cannot, however, be proved presently.
\section{Conclusion}
%%%%%%%%%%%%%%%%%%%%%%%%%%%%%%%%%%%%%%%%%%%%%%%%
In this paper, we studied time-dependent Abelian and non-Abelian AB effects. We showed that for a superconductor exposed to a time-varying magnetic field, there is no constraint on the magnetic flux due to the presence of zero phase factor in both Abelian and non-Abelian  AB effects. We also investigated the non-Abelian AB effect and proved
that, for  time-dependent non-Abelian magnetic fields, the AB phase disappears up to the first order in the expansion of the phase factor.

\end{document}